\documentclass[10pt,twocolumn,twoside]{eec_style}
\usepackage{epsfig}
\usepackage{subfloat}
\usepackage{array}
\usepackage{setspace}
\usepackage{enumerate}
\usepackage{amsbsy,amsmath,subfigure}
\usepackage{amssymb,subfigure}
\graphicspath{{../jpeg/eps/jp2}}
\usepackage{graphicx}
\usepackage{eurosym}
\usepackage{multirow}
\usepackage{algorithmic}
\usepackage{algorithm}
\usepackage{mathtools}
\usepackage{color}
\usepackage[table]{xcolor}
\usepackage{url}
\usepackage{bm}
\usepackage{cite}
\begin{document}
\newcommand{\eqn}[1]{(\ref{eqn:#1})}
\newcommand{\rank}{{\rm rank\;}}
\newcommand{\diag}{{\rm diag\;}}
\newcommand{\fig}[1]{Fig. \ref{fig:#1}}
\newcommand{\refsec}[1]{Section \ref{sec:#1}\hspace{0.2cm}}
\newtheorem{theorem}{Theorem}
\newtheorem{lemma}{Lemma}
\newtheorem{remarks}{Remarks}
\newtheorem{note}{Note}
\newtheorem{definition}{Definition}
\DeclareGraphicsExtensions{.jpg,.pdf,.mps,.png,.eps}
\title{Block-Wise Encryption for Reliable Vision Transformer models}
\authorlist{%
 \authorentry{Hitoshi Kiya\( ^{1} \)}{n}{}
 \authorentry{Ryota Iijima\( ^{2} \)}{n}{}
 \authorentry{Teru Nagamori\( ^{3} \)}{n}{}
}
\received{2014}{9}{3}{2014}{11}{27}
\runninghead{Ecti Transactions on Electrical Eng., Electronics, and Communications}
%\thanks{A part of this work ....}
\setcounter{page}{1}

\maketitle
\begin{abstract}
This article presents block-wise image encryption for the vision transformer and its applications.
Perceptual image encryption for deep learning enables us not only to protect the visual information of plain images but to also embed unique features controlled with a key into images and models.
However, when using conventional perceptual encryption methods, the performance of models is degraded due to the influence of encryption.
In this paper, we focus on block-wise encryption for the vision transformer, and we introduce three applications: privacy-preserving image classification, access control, and the combined use of federated learning and encrypted images.
Our scheme can have the same performance as models without any encryption, and it does not
require any network modification. It also allows us to easily update the secret key.
In experiments, the effectiveness of the scheme is demonstrated in terms of performance degradation and access control on the CIFAR-10 and CIFAR-100 datasets.

\end{abstract}


\footnotetext{\hspace{2mm}Final manuscript received on June 6, 2017.}
\footnotetext{\hspace{2mm}\( ^{1,2} \) The authors are with university and address, E-mail:}

% Introduction

\section{Introduction}
Deep neural networks (DNNs) have been deployed in many applications including security critical ones such as biometric authentication, automated driving, and medical image analysis \cite{lecun2015deep, liu2019recent}.
Training successful models also requires three ingredients: a huge amount of data, GPU accelerated computing resources, and efficient algorithms, and it is not a trivial task.
In fact, collecting images and labeling them is also costly and will also consume a massive amount of resources.
Therefore, trained ML models have great business value.
Considering the expenses necessary for the expertise, money, and time taken to train a model, a model should be regarded as a kind of intellectual property (IP).
In addition, generally, data contains sensitive information, and it is difficult to train a model while preserving privacy.
In particular, data with sensitive information cannot be transferred to untrusted third-party cloud environments (cloud GPUs and TPUs) even though they provide a powerful computing environment \cite{huang2014survey, moo2013p3, lagendijk2013encrypted, fredrikson2015model, shokri2017membership, christian2014intriguing, siva2020adversarial}.
Accordingly, it has been challenging to train/test a DNN model with encrypted images as one way for solving these issues \cite{kiya2022overview}. 
However, when using conventional perceptual encryption methods, the performance of models is degraded due to the influence of encryption.\par
In this paper, we present a block-wise encryption method for achieving reliable vision transformer (ViT) models.
In the method, a model trained with plain images is transformed with a secret key to give unique features controlled with the key to the model, and encrypted images are applied to the model.
In addition, three applications: privacy-preserving image classification, access control, and the combined use of federated learning and encrypted images, are presented to show the effectiveness of our method.
In the method, the vision transformer (ViT) \cite{Alexey2021an}, which is known to have a high performance, is used to reduce the influence of block-wise encryption thanks to its architecture.
It allows us not only to obtain the same performance as models trained with plain images but to also update the secret key easily.
In experiments, our method is evaluated in terms of performance degradation and access control in an image classification task on the CIFAR-10 and CIFAR-100 datasets.

\section{Related Work}
Image encryption methods for deep learning and ViT are summarized here.

\subsection{Image Encryption for Deep Learning}
Various image transformation methods with a secret key, often referred to as perceptual image encryption or image cryptography, have been studied so far for many applications.
Figure 1 shows typical applications of image encryption with a key.
Image encryption with a key allows us not only to protect the visual information of plain images but to also embed unique features controlled with the key into images.
The use of visually protected images has enabled various kinds of applications. \par
One of the origins of image transformation with a key is in block-wise image encryption schemes for encryption-then-compression (EtC) systems \cite{chuman2019encryption, chuman2017security, zhou2014designing, ghonge2014a, liu2018ecg, liu2010efficient, hu2014a, johnson2004on, methaq2016an, grayscale2019warit}.
Image encryption prior to image compression is required in certain practical scenarios such as secure image transmission through an untrusted channel provider.
An EtC system is used in such scenarios, although the traditional way of securely transmitting images is to use a compression-then-encryption (CtE) system.
Compressible encryption methods have been applied to privacy-preserving compression, data hiding, and image retrieval \cite{Shoko2020a, iida2020privacy, iida2019an} in cloud environments.
In addition, visually protected images have been demonstrated to be effective in privacy-preserving learning \cite{kiya2022overview, kawamura2020aprivacy, bandoh2020distributed, takayuki2020secure, nakachi2020privacy, ibuki2016unitary, maekawa2019privacy},
adversarial defense \cite{aprilpyone2021block, maung2020encryption, maung2021ensemble}, access control \cite{chen2018protect, chen2019deepattest, maungmaung2021a}, and DNN
watermarking \cite{maung2021ensemble, uchida2017embedding, chen2019deepmarks, rouhani2018deepsigns, fan2021deepip, adi2018turning, zhang2018protecting, sakazawa2019visual, le2019adversarial}.  \par
In this paper, we focus on image encryption for deep learning, called learnable encryption, under the use of ViT.
In addition, it is demonstrated to be useful to privacy-preserving classification, access control, and federated learning with encrypted images while maintaining the high performance that ViT has.

\begin{figure*}[t]
    \centering
    \includegraphics[scale=0.3]{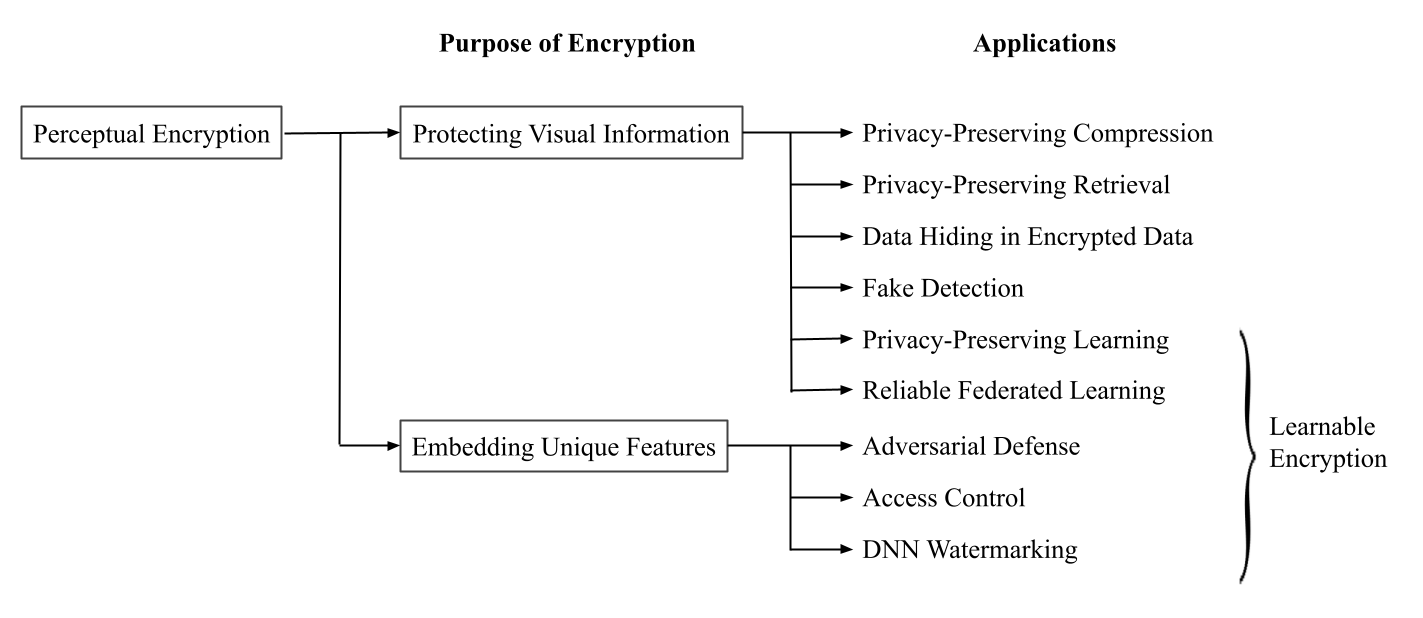}
    \caption{Applications of perceptual image encryption}
    \label{fig:1}
\end{figure*}

\subsection{Vision Transformer}
The transformer architecture has been widely used in natural language processing (NLP) tasks \cite{devlin2019bert}.
The vision transformer (ViT) \cite{Alexey2021an} has also provided excellent results compared with state-of-the-art convolutional networks.
Following the success of ViT, several isotropic networks (with the same depth and resolution across different layers in the network) have been proposed such as MLP-Mixer \cite{ilya2021mlpmixer}, ResMLP \cite{hugo2021resmlp}, CycleMLP \cite{chen2022cyclemlp}, gMLP \cite{liu2021pay}, vision permutator \cite{Hou2022VisionPA}, and ConvMixer \cite{trockman2022patches}. \par

Figure 2 illustrates the architecture of ViT, where ViT consists of two embedding processes (patch embedding and position embedding) and a transformer encoder.
In ViT, an input image $x \in {\mathbb{R}}^{h \times w \times c}$ is segmented into $N$ patches with a size of $ p \times p $, where $h$, $w$, and $c$ are the height, width, and number of channels of the image.
In addition, an integer $N$ is given as $hw / p^2$.
After that, each patch is flattened as $x^{i}_{\mathrm{p}} = [x^{i}_{\mathrm{p}}(1), x^{i}_{\mathrm{p}}(2), \dots, x^{i}_{\mathrm{p}}(L)]$, where $L = p^{2} c$.
Finally, a sequence of embedded patches is given as

\begin{equation}
    z_{0} = [x_{\mathrm{class}}; x_{\mathrm{p}}^{1} {\mathbf{E}}; x_{\mathrm{p}}^{2} {\mathbf{E}}; \dots x_{\mathrm{p}}^{i} {\mathbf{E}}; \dots x_{\mathrm{p}}^{N } {\mathbf{E}}] + {\mathbf{E_{\mathrm{pos}}}},
\end{equation}

\noindent
where

\begin{equation}
    \begin{aligned}
        & \mathbf{E_{\mathrm{pos}}} = ((e_{\mathrm{pos}}^0)^{\top} (e_{\mathrm{pos}}^1)^{\top} \dots (e_{\mathrm{pos}}^i)^{\top} \dots (e_{\mathrm{pos}}^N)^{\top})^{\top},\\
        & x_{\mathrm{class}} \in {\mathbb{R}}^{D}, x^{i}_{\mathrm{p}} \in {\mathbb{R}}^{L}, e^{i}_{\mathrm{pos}} \in {\mathbb{R}}^{D}, \\
        & \mathbf{E} \in {\mathbb{R}}^{L \times D}, \ \mathbf{E_{\mathrm{pos}}} \in {\mathbb{R}}^{(N+1) \times D}. \nonumber
    \end{aligned}
\end{equation}

\noindent
$x_{\mathrm{class}}$ is the classification token, $\mathbf{E}$ is the embedding (patch embedding) to linearly map each patch to dimensions $D$, $\mathbf{E}_{\mathrm{pos}}$ is the embedding (position embedding) that gives position information to patches in the image, $e^{0}_{\mathrm{pos}}$ is the information of the classification token, and $e^{i}_{\mathrm{pos}}, \: i = 1, \dots, N$, is the position information of each patch. \par
In patch embedding, patches are mapped to vectors, and the position information is embedded in position embedding.
In this paper, we encrypt not only test images but also two embeddings: patch embedding $\mathbf{E}$ and position embedding $\mathbf{E}_{\mathrm{pos}}$, in a trained model.
The resulting sequence of vectors is fed to a standard transformer encoder, and the output of the transformer is provided to a multi-layer perceptron (MLP) to get an estimation result.

\begin{figure}[t]
    \centering
    \includegraphics[scale=0.31]{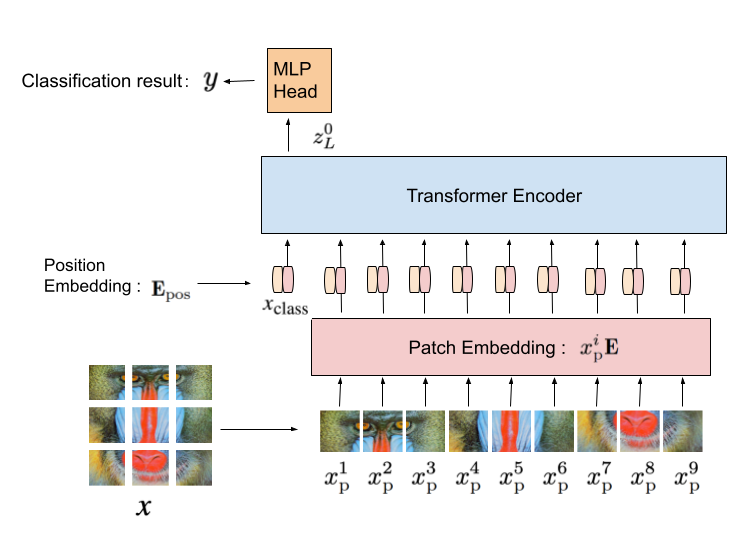}
    \caption{Architecture of ViT \cite{Alexey2021an}}
    \label{fig:2}
\end{figure}

\section{Image Encryption for Vision Transformer}
An encryption method with random numbers is presented here.
The method makes it possible to avoid the performance degradation of models even when using encrypted images.

\subsection{Overview}
Figure 3 shows the scenario of the presented scheme in a privacy-preserving image classification task, where it is assumed that the model builder is trusted, and the service provider is untrusted.
The model builder trains a model by using plain images and encrypts the trained model with a key $\mathrm{K}$. \par
The encrypted model is given to the service provider, and the key is sent to a client.
The client prepares a test image encrypted with the key and sends it to the service provider.
The encrypted test image is applied to the encrypted model to obtain an estimation result, and the result is sent back to the client.
Note that the provider has neither a key nor plain images. The framework presented in this paper enables us to achieve this scenario without any performance degradation compared with the use of plain images.

\begin{figure*}[t]
    \centering
    \includegraphics[scale=0.35]{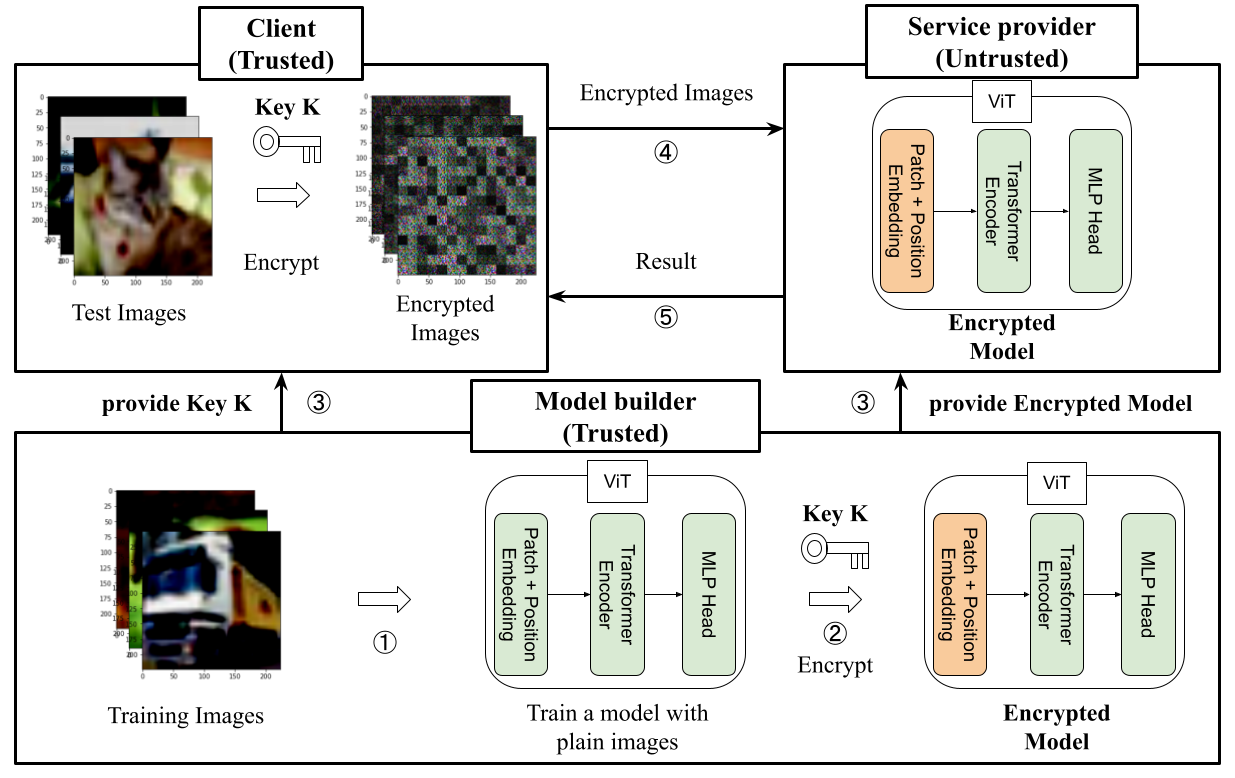}
    \caption{ Scenario of proposed scheme }
    \label{fig:3}
\end{figure*}

\subsection{Model Encryption}
As shown in Figure 3, a trained model is encrypted by using random numbers (key).
In the method, patch embedding E and position embedding $\mathbf{E}_{\mathrm{pos}}$ in Eq. (1) are encrypted by using random matrices, respectively.

\subsubsection{Patch Embedding Encryption}
The following transformation matrix $\mathbf{E_{a}}$ is used to encrypt patch embedding $\mathbf{E}$.

\begin{gather}
    \mathbf{E_{a}} = 
    \begin{bmatrix}
        k_{(1,1)} & k_{(1,2)} & \dots & k_{(1,L)} \\
        k_{(2,1)} & k_{(2,2)} & \dots & k_{(2,L)} \\
        \vdots & \vdots & k_{(i,j)} & \vdots \\
        k_{(L,1)} & k_{(L,2)} & \dots & k_{(L,L)}
    \end{bmatrix},
\end{gather}
where
\begin{gather}
    {\mathbf{E_{a}}} \in {\mathbb{R}}^{L \times L}, \: {\mathrm{det}} \: {\mathbf{E_{a}}} \neq 0, \nonumber \\
    k_{(i,j)} \in {\mathbb{R}}, \: i,j \in \{ 1, \dots, L \}. \nonumber
\end{gather}

\noindent
Note that the element values of $\mathbf{E_{a}}$ are randomly decided, but $\mathbf{E_{a}}$ has to have an inverse matrix. \par
Then, by multiplying $\mathbf{E}$ by $\mathbf{E_{a}}$, an encrypted patch embedding $\mathbf{\hat{E}}$ is given as

\begin{equation}
    \mathbf{\hat{E}} = \mathbf{E_{a}} \mathbf{E}.
\end{equation}

\subsubsection{Position Embedding Encryption}
Position embedding ${\mathbf{E}}_{\mathrm{pos}}$ is encrypted as below.
\begin{enumerate}
    \setlength{\leftskip}{0.5cm}
    \renewcommand{\labelenumi}{\arabic{enumi}).}
    \item Generate a random integer vector with a length of $N$ as
    \begin{equation}
        l_{t} = [l_{e}(1), l_{e}(2), \dots, l_{e}(i), \dots, l_{e}(N)],
    \end{equation}
    where
    \begin{gather}
        l_{e}(i) \in \{ 1, 2, \dots, N \}, \nonumber \\
        l_{e}(i) \neq l_{e}(j) \: \: {\mathrm{if}} \: \: i \neq j, \nonumber \\
        i,j \in \{ 1, \dots, N \}. \nonumber
    \end{gather}
    \item Calculate $m_{(i,j)}$ as
    \begin{equation}
        m_{(i,j)} = \left \{
        \begin{aligned}
        & 0 && \: (j \neq l_{e}(i)) \\
        & 1 && \: (j = l_{e}(i)).
        \end{aligned}
    \right .
    \end{equation}
    \item Define a random matrix as
    \begin{gather}
        \mathbf{E_{b}} =
        \begin{bmatrix}
            1 & 0 & 0 & \dots & 0 \\
            0 & m_{(1,1)} & m_{(1,2)} & \dots & m_{(1,N)} \\
            0 & m_{(2,1)} & m_{(2,2)} & \dots & m_{(2,N)} \\
            \vdots & \vdots & \vdots & \ddots & \vdots \\
            0 & m_{(N,1)} & m_{(N,2)} & \dots & m_{(N,N)}
        \end{bmatrix},
    \end{gather}
    where
    \begin{gather}
        {\mathbf{E_{b}}} \in {\mathbb{R}}^{(N+1) \times (N+1)}. \nonumber
    \end{gather}
    For instance, if $N = 3$ and $l_{t} = [1,3,2]$, $\mathbf{E_{b}}$ is given by
    \begin{gather}
        \mathbf{E_{b}} =
        \begin{bmatrix}
            1 & 0 & 0 & 0 \\
            0 & 1 & 0 & 0 \\
            0 & 0 & 0 & 1 \\
            0 & 0 & 1 & 0
        \end{bmatrix}.
    \end{gather}
    \item  Transform $\mathbf{E}_{\mathrm{pos}}$ to $\mathbf{\hat{E}}_{\mathrm{pos}}$ as
    \begin{gather}
        \mathbf{\hat{E}}_{\mathrm{pos}} = \mathbf{E_{b}} {\mathbf{E}}_{\mathrm{pos}}.
    \end{gather}
\end{enumerate}

\subsection{Test Image Encryption}
A test image $x \in {\mathbb{R}}^{h \times w \times c}$ is transformed into an encrypted image $\tilde{x} \in {\mathbb{R}}^{h \times w \times c}$ as below (see Figure 4).
\begin{enumerate}
    \setlength{\leftskip}{0.5cm}
    \renewcommand{\labelenumi}{(\alph{enumi})}
    \item Divide $x$ into $N$ non-overlapped blocks with a size of $p \times p$ such that $B = \{ B_{1}, \dots, B_{N} \}$, where $p \times p$ is the same size as the patch size used in a ViT model.
    \item Generate permutated blocks $\bar{B}$ by
    \begin{equation}
        \begin{aligned}
            \bar{B} &= {\mathbf{E_{b}}} B \\
            &= \{ \bar{B}_{1}, \dots, \bar{B}_{N} \},
        \end{aligned}
    \end{equation}
    where $B \in {\mathbb{R}}^{1 \times N}$.
    \item Flatten each block $\bar{B}_{i} \in {\mathbb{R}}^{p \times p \times c}$ into a vector $\bar{x}^{i}_{\mathrm{p}} \in {\mathbb{R}}^{p^{2} c}$ as
    \begin{gather}
        \bar{x}^{i}_{\mathrm{p}} = [\bar{x}^{i}_{\mathrm{p}}(1), \dots, \bar{x}^{i}_{\mathrm{p}}(L)].
    \end{gather}
    Note that the following relation is satisfied.
    \begin{equation}
        \begin{aligned}
            & [x_{\mathrm{class}}; {\bar{x}}^{1}_{\mathrm{p}}; \dots, {\bar{x}}^{i}_{\mathrm{p}}; \dots; {\bar{x}}^{N}_{\mathrm{p}}] \\
            = & {\mathbf{E_{b}}} [x_{\mathrm{class}}; x^{1}_{\mathrm{p}}; \dots, x^{i}_{\mathrm{p}}; \dots; x^{N}_{\mathrm{p}}]
        \end{aligned}
    \end{equation}
    % \begin{gather}
    %     \bar{x},x,E_b
    %     [{\bar{x}}^{i}_{\mathrm{p}}, \dots, {\bar{x}}^{i}_{\mathrm{p}}]
    % \end{gather}
    \item Generate an encrypted vector $\tilde{x}^{i}_{\mathrm{p}}$ by multiplying vector $\bar{x}^{i}_{\mathrm{p}}$ by matrix ${\mathbf{E_{a}}}^{-1} \in {\mathbb{R}}^{L \times L}$ as
    \begin{gather}
        \tilde{x}^{i}_{\mathrm{p}} = \bar{x}^{i}_{\mathrm{p}} {\mathbf{E_{a}}}^{-1}.
    \end{gather}
    \item Rebuild vector $\tilde{x}^{i}_{\mathrm{p}}$ into block $\tilde{B^i}$ in the reverse order of step (c).
    \item Concatenate $\tilde{B} = \{ \tilde{B}^{1}, \dots \tilde{B}^{N} \}$ into an encrypted test image $\tilde{x}$.
\end{enumerate}
\noindent
Figure 5 shows an example of images encrypted with this procedure. \par
When replacing $\mathbf{E}$, ${\mathbf{E}}_{\mathrm{pos}}$, and $x^{i}_{\mathrm{p}}$ with $\mathbf{\hat{E}}$, $\mathbf{\hat{E}}_{\mathrm{pos}}$, and $\tilde{x}^{i}_{\mathrm{p}}$, respectively, the sequence in Eq. (1) is reduced to
\begin{gather}
    \tilde{z}_{0} = [x_{\mathrm{class}};, \tilde{x}^{1}_{\mathrm{p}} {\mathbf{\hat{E}}}; \dots; \tilde{x}^{i}_{\mathrm{p}} {\mathbf{\hat{E}}}; \dots; \tilde{x}^{N}_{\mathrm{p}} {\mathbf{\hat{E}}}] + \mathbf{\hat{E}}_{\mathrm{pos}}.
\end{gather}
Thus, by substituting Eqs. (3),(8), and (11) with Eq. (13), we obtain:
\begin{align}
    \begin{split}
        \tilde{z}_{0} = {} & [x_{\mathrm{class}}; \bar{x}^{1}_{\mathrm{p}} {\mathbf{E_{a}}}^{-1} {\mathbf{E_{a}}} {\mathbf{E}}; \dots; \bar{x}^{i}_{\mathrm{p}} {\mathbf{E_{a}}}^{-1} {\mathbf{E_{a}}} {\mathbf{E}}; \dots; \\
        & \bar{x}^{N}_{\mathrm{p}} {\mathbf{E_{a}}}^{-1} {\mathbf{E_{a}}} {\mathbf{E}}] + {\mathbf{E_{b}}} {\mathbf{E}}_{\mathrm{pos}}
    \end{split} \nonumber
    \\
    = {} & {\mathbf{E_{b}}} [x_{\mathrm{class}}; x^{1}_{\mathrm{p}} {\mathbf{E}}; \dots; x^{i}_{\mathrm{p}} {\mathbf{E}}; \dots; x^{N}_{\mathrm{p}} {\mathbf{E}}] + {\mathbf{E_{b}}}{\mathbf{E}}_{\mathrm{pos}} \nonumber \\
    = {} & {\mathbf{E_{b}}} z_{0}
\end{align}
From the equation, the influence of encryption can be avoided except for ${\mathbf{E_{b}}}$.
Accordingly, the encrypted model allows us to have the same performance as that of the model trained with plain images, if test images are encrypted with the same key as that used for model encryption.

\begin{figure*}[t]
    \centering
    \includegraphics[scale=0.32]{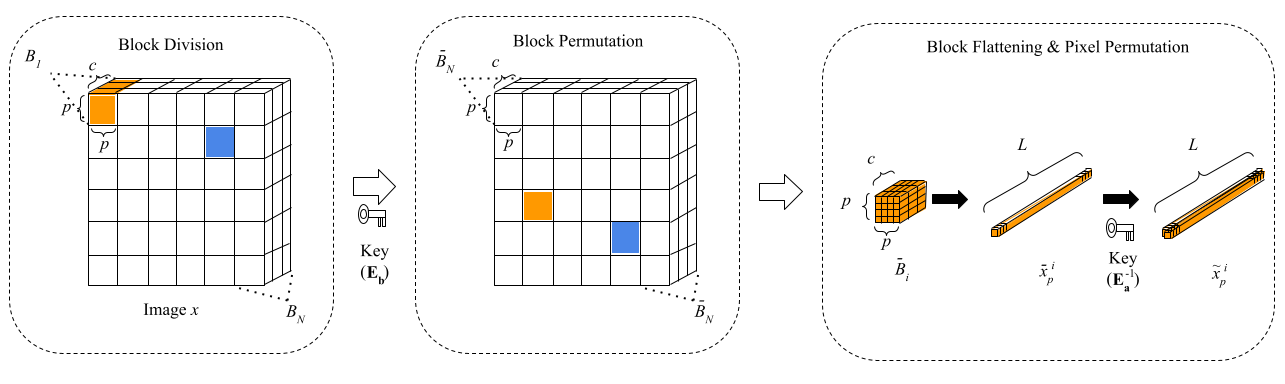}
    \caption{Procedure of block-wise encryption}
    \label{fig:4}
\end{figure*}

\subsection{Generation of Random Matrices}
A random orthogonal matrix can be generated as a random matrix $\mathbf{E_{a}}$ by using Gram–Schmidt orthonormalization \cite{ibuki2016unitary,wang2007face}.
The procedure for generating $\mathbf{E_{a}}$ with a size of $L \times L$ is given as follows.
\begin{enumerate}
    \setlength{\leftskip}{0.5cm}
    \item Generate a real matrix $\mathbf{R}$ with a size of $L \times L$ by using a random number generator with a seed.
    \item Calculate $\mathrm{det}({{\mathbf{R}}})$, and proceed to step 3 if $\mathrm{det}({{\mathbf{R}}}) \neq 0$.
    Otherwise, return to step 1.
    \item Compute a random orthogonal matrix $\mathbf{E_{a}}$ from $\mathbf{R}$ by using Gram-Schmidt orthogonalization.
\end{enumerate}
In this framework, any regular matrix can be used as $\mathbf{E_{a}}$ for image encryption.
Several conventional methods for privacy-preserving image classification use permutation matrices of pixel values \cite{hitoshi2023image,kiya2022privacy}, in which many elements have zero values in matrices as
\begin{gather}
    {\mathbf{E_{a}}} =
    \begin{bmatrix}
        0 & 1 & 0 \\
        0 & 0 & 1 \\
        1 & 0 & 0
    \end{bmatrix}.
\end{gather}
In contrast, the random orthogonal matrices generated with Gram-Schmidt orthogonalization include no zero values as elements in general.
The use of such matrices allows us not only to more strongly protect the visual information of plain images but to also enhance robustness against various attacks while maintaining the same performance as that of models trained with plain images.
In addition, ${\mathbf{E_{a}}}^{-1}$ can easily be calculated as the transposed matrix of $\mathbf{E_{a}}$.

\subsection{Properties of Proposed Scheme}
When the block size for image encryption is the same as the patch size used in a ViT model.
The proposed method has the following properties.
\begin{itemize}
    \item The model performs well only if test images are transformed with the same key as that used for transforming the model from Eq. (14).
    \item The method does not cause any performance degradation in terms of the accuracy of models as in Eq. (14).
    \item Model training and encryption are independent (see Figure 3). Therefore, it is possible to easily update a key.
\end{itemize}

\section{Applications of Encrypted ViT}
Three applications of encrypted ViT models are presented to demonstrate the
usefulness of the encryption scheme.

\subsection{Privacy-preserving Image Classification}
One of the applications is to use encrypted ViT models for privacy-preserving image classification as shown in Figure 3, in which visually protected test images are sent to an untrusted provider. \par
A threat model includes a set of assumptions such as attacker's goals, knowledge, and capabilities.
Users without secret key $K$ are assumed to be the adversary.
In this application, we consider the attacker's goal to be to restore visual information from encrypted test images.
We assume that authorized users know key $K$, and the model owner securely manages both key $K$ and the trained model without any encryption.
In addition, the encryption method is also assumed to be disclosed except for key $K$.
Thus, an adversary may perform ciphertext-only (COA) attacks via this information to restore the perceptual information from encrypted images. \par
Accordingly, the encryption method should satisfy the following requirements.
\begin{enumerate}
    \setlength{\leftskip}{0.5cm}
    \item Security: No perceptual information of plain images should be reconstructed from encrypted images unless the key is exposed.
    \item Model capability: Privacy-preserving methods for DNNs should maintain an approximate accuracy as when using plain images.
    \item Computational requirement: Privacy-preserving DNNs should not increase the computational requirement in quantity.
    \item Key update: The key should easily be updated without re-training the model.
\end{enumerate}
In experiments, the effectiveness of our scheme will be evaluated in terms of the above requirements.

\subsection{Access Control with Encypted Model}
The second application is to protect a model from misuse when it has been stolen, referred to as access control that aims to protect the functionality of DNN models from unauthorized access.
Trained models have great business value. Considering the expenses necessary for the expertise, money, and time taken to train a model, trained models should be regarded as a kind of intellectual property (IP). \par
Accordingly, encrypted models are required not only to provide high performance to authorized users but also low performance to unauthorized users.
Our scheme is effective in access control in addition to privacy-preserving deep learning as verified in an experiment.

\subsection{Federated Learning in Combination with Encryption}
It has been very popular for data owners to train and test deep neural network (DNN) models in cloud environments. 
However, data privacy such as personal medical records may be compromised in cloud environments, so privacy-preserving methods for deep learning have become an urgent problem. \par
One of the solutions is to use federated   learning (FL) \cite{jakub2016federated,McMahan2017communication},   which was proposed by Google.
FL is capable of significantly preserving clients' private data from being exposed to adversaries.
However, FL aims to construct models over multiple participants without directly sharing their raw data, so the privacy of test (query) images is not considered. \par
Another approach is to encrypt a trained model, and then encrypted test (query) images are applied to the encrypted model shown in Figure 3.
However, this approach does not consider constructing models over multiple participants without directly sharing their raw data, although the visual information of test images can be protected. \par
For these reasons, the combined use of FL  and encrypted test images is effective in privacy-preserving image classification tasks with ViT \cite{kiya2022overview} (see Figure 6).
The method allows us not only to train models over multiple participants without directly sharing their raw data but to also protect the privacy of test (query) images.
In addition, it can maintain the same accuracy as that of models normally trained with plain images.

\begin{figure}
    \begin{tabular}{ccc}
        \begin{minipage}[b]{0.26\linewidth}
            \centering
            \includegraphics[keepaspectratio, scale=0.2]{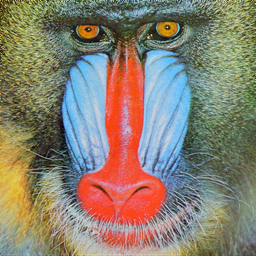}
        \end{minipage}&
        \begin{minipage}[b]{0.26\linewidth}
            \centering
            \includegraphics[keepaspectratio, scale=0.1]{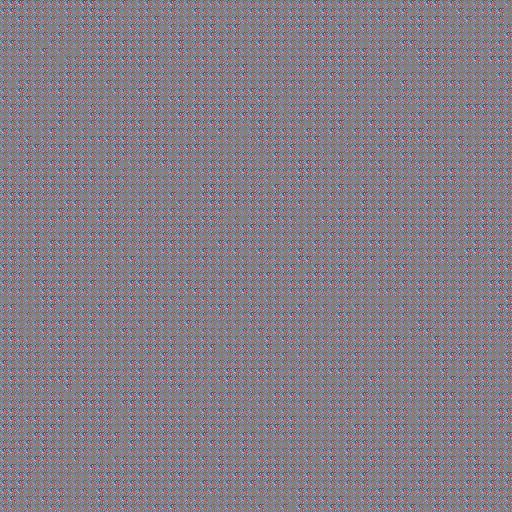}
        \end{minipage}&
        \begin{minipage}[b]{0.26\linewidth}
            \centering
            \includegraphics[keepaspectratio, scale=0.1]{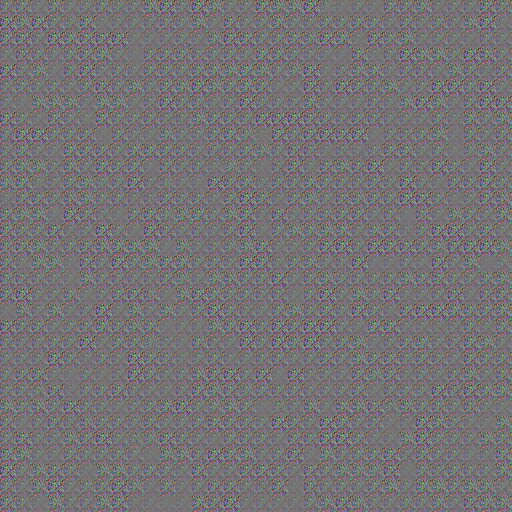}
        \end{minipage}\\
        \begin{minipage}[b]{0.26\linewidth}
            \centering
            \includegraphics[keepaspectratio, scale=0.2]{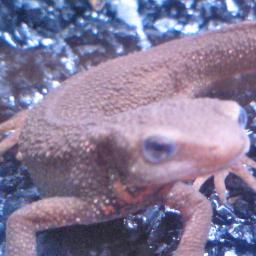}
        \end{minipage}&
        \begin{minipage}[b]{0.26\linewidth}
            \centering
            \includegraphics[keepaspectratio, scale=0.1]{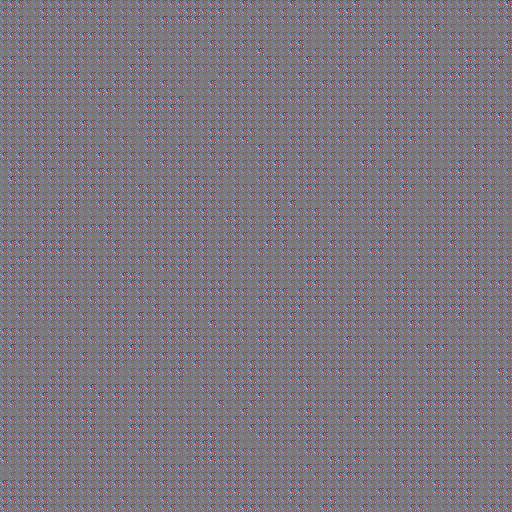}
        \end{minipage}&
        \begin{minipage}[b]{0.26\linewidth}
            \centering
            \includegraphics[keepaspectratio, scale=0.1]{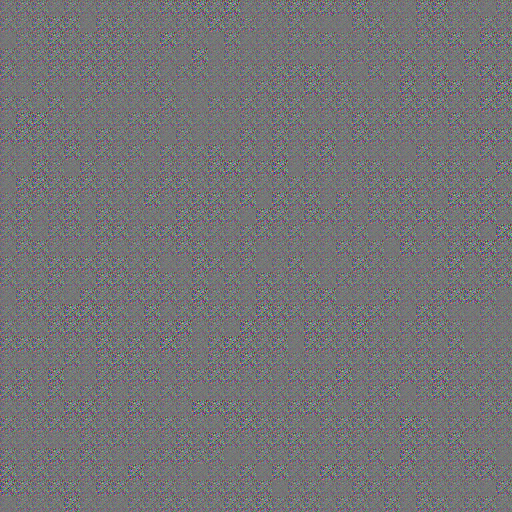}
        \end{minipage} \\
        Plain image & $p=8$ & $p=16$
    \end{tabular}
    \caption{Example of encrypted images}
\end{figure}

\begin{figure}[t]
    \centering
    \includegraphics[scale=0.2]{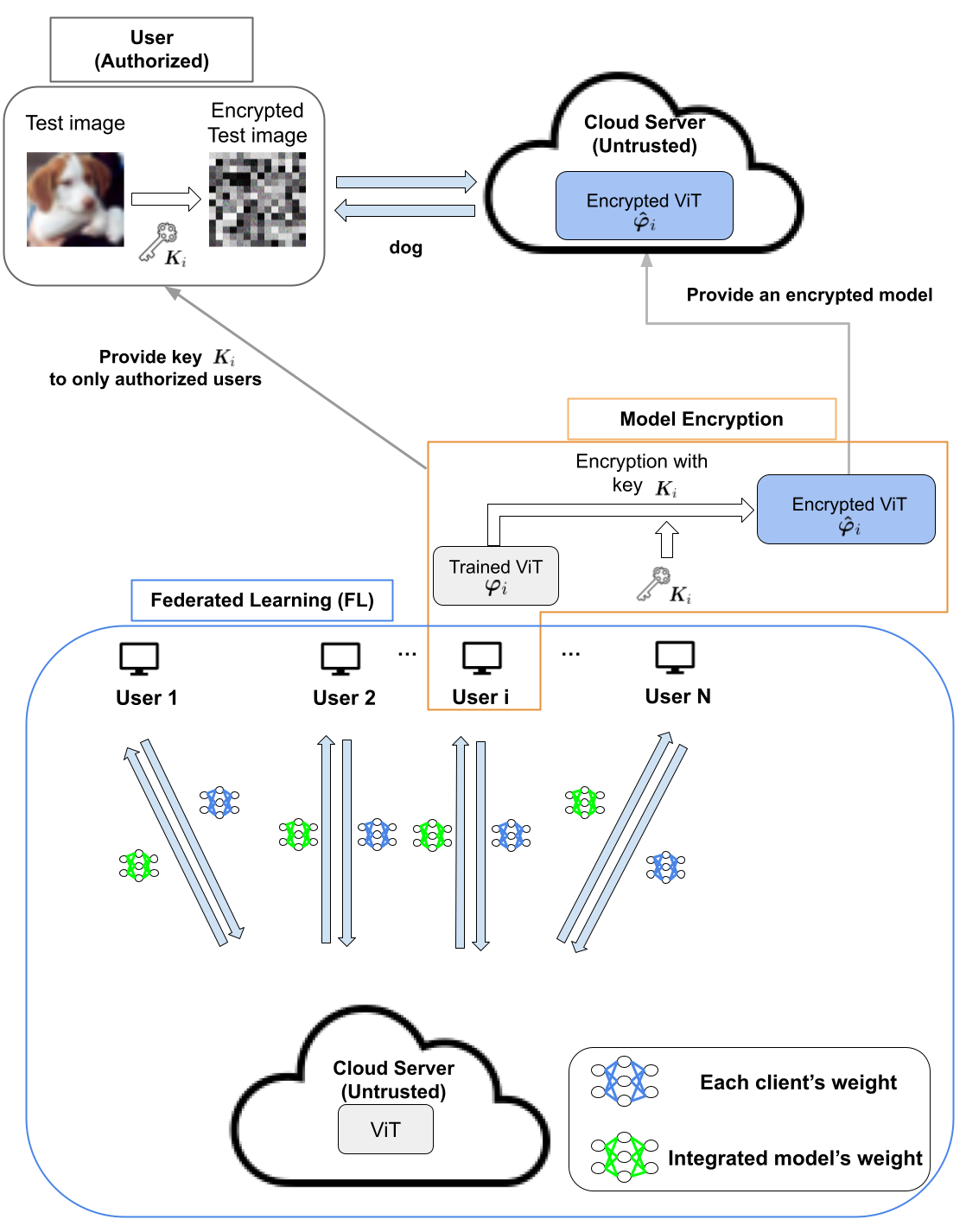}
    \caption{Combined use of FL and encrypted test images}
    \label{fig:5}
\end{figure}

\section{Experiment and Discussion}
In experiments, the effectiveness of encrypted ViT models is verified in an image classification task.

\subsection{Privacy Preservation}
As the first application of encrypted ViT models, a privacy-preserving image classification task was carried out in accordance with the framework in 4.1 where visual information on test images is protected \cite{hitoshi2023image}.

\subsubsection{Experiment Setup}
To confirm the effectiveness of the presented scheme, experiments were carried on the CIFAR-10 dataset (with $10$ classes).
The dataset consists of 60,000 color images (dimension of $3 \times 32 \times 32$), where $50,000$ images are for training, $10,000$ for testing, and each class contains $6,000$ images.
Images in the dataset were resized to $3 \times 224 \times 224$ to input them to ViT, before applying the proposed encryption algorithm, where the block size was $16 \times 16$.
We used the PyTorch \cite{paszke2019pytorch} implementation of ViT and fine-tuned a ViT model with a patch size $P=16$, which was pre-trained on ImageNet-21k.
The ViT model was fine-tuned for $5000$ epochs.
The parameters of the stochastic  gradient descent (SGD) optimizer were a momentum of $0.9$ and a learning rate  value of $0.03$. \par
In addition, we used three conventional visual information protection methods (Tanaka's method \cite{tanaka2018learnable}, the pixel-based encryption method \cite{sirichotedumrong2019privacy}, and the GAN-based transformation method \cite{sirichotedumrong2021a}) to compare them with our method.
ResNet-20 was used to validate the effectiveness of the conventional method with reference to \cite{ito2021image}.
CIFAR-10 was also used for training networks, and the networks were trained for $200$ epochs by using SGD with a weight decay of $0.0005$ and a momentum of $0.9$.
The learning rate was initially set to $0.1$, and it was multiplied by $0.2$ at $60$, $120$, and $160$ epochs. The batch size was $128$.

\subsubsection{Performance of Encrypted ViT}
First, we compared the proposed method with conventional ones in terms of the accuracy of image classification under the use of ViT and ResNet-20.
As shown in Table 1, the performance of all conventional methods was degraded compared with the baselines, which were results calculated with plain images.
In contrast, the proposed method did not degrade the performance at all.
Accordingly, our method was verified to be able to maintain the same accuracy as that of the baselines as shown in Eq. (14). 

% \begin{figure}
%     \begin{tabular}{ccc}
%         \begin{minipage}[b]{0.26\linewidth}
%             \centering
%             \includegraphics[keepaspectratio, scale=0.2]{figure/Mandrill.png}
%         \end{minipage}&
%         \begin{minipage}[b]{0.26\linewidth}
%             \centering
%             \includegraphics[keepaspectratio, scale=0.1]{figure/Mandrill_512_OM8_bs.png}
%         \end{minipage}&
%         \begin{minipage}[b]{0.26\linewidth}
%             \centering
%             \includegraphics[keepaspectratio, scale=0.1]{figure/Mandrill_512_OM16_bs.png}
%         \end{minipage}\\
%         \begin{minipage}[b]{0.26\linewidth}
%             \centering
%             \includegraphics[keepaspectratio, scale=0.2]{figure/imori.jpg}
%         \end{minipage}&
%         \begin{minipage}[b]{0.26\linewidth}
%             \centering
%             \includegraphics[keepaspectratio, scale=0.1]{figure/imori_512_OM8_bs.png}
%         \end{minipage}&
%         \begin{minipage}[b]{0.26\linewidth}
%             \centering
%             \includegraphics[keepaspectratio, scale=0.1]{figure/imori_512_OM16_bs.png}
%         \end{minipage} \\
%         Plain image & $p=8$ & $p=16$
%     \end{tabular}
%     \caption{Example of encrypted images}
% \end{figure}

\begin{table}[t]
    \centering
    \caption{Comparison with conventional methods in terms of classification accuracy}
    \begin{tabular}{c|c|c}
    \hline
    Model & Method & Accuracy \\
    \hline
    \hline
    ViT & Baseline & 99.03 \\
    & Ours & \textbf{99.03} \\
    \hline
    ResNet-20 \cite{ito2021image} & Baseline & 91.55 \\
    & Tanaka \cite{tanaka2018learnable} & 87.02 \\
    & Pixel-based \cite{sirichotedumrong2019privacy} & 86.66 \\
    & GAN-based \cite{sirichotedumrong2021a} & 82.55 \\
    \hline
    \end{tabular}
    \label{tab:1}
\end{table}

\subsubsection{Visual Protection}
Figure 5 shows an example of images encrypted with the method in 3.3, where random matrices $\mathbf{E_{a}}$ were generated by using Gram-Schmidt orthonormalization.
The images had $H \times W \times C = 512 \times 512 \times 3$ as an image size, and the block sizes used for encryption were $p = 8$ and $p = 16$.
From the figures, the encrypted images have almost none of the visual information of the plain images. \par
In addition to visual protection, encrypted images have to be robust enough against various attacks, which aim to restore visual information from encrypted images.
ViT has two embeddings: position embedding and patch embedding, so not only pixel values in every block but also the position of blocks can be changed randomly.
We already confirmed that the encryption including block permutation is robust against cipher-text-only attacks (COAs) including jigsaw puzzle solver attacks \cite{chuman2023ajigsaw}.
In particular, the use of random matrices generated with Gram-Schmidt orthonormalization is more robust than that of simple permutation matrixes.

\subsection{Application to Access Control}
Next, we validated whether our method could protect models where the experimental conditions were the same as those in 5.1.1 \cite{hitoshi2023image}.
Table 2 shows the accuracy of image classification when encrypted or plain images were input to the encrypted model.
The encrypted model performed well for test images with the correct key, but its accuracy was not high when using plain test images.
The CIFAR-10 dataset consists of ten classes, so $9.06$ is almost the same accuracy as that when test images are randomly classified. \par
Next, we confirmed the performance of images encrypted with a different key from that used in the model encryption.
We prepared 100 random keys, and test images encrypted with the keys were input to the encrypted model.
From the box plot in Figure 7, the accuracy of the models was not high under the use of
the wrong keys.
Accordingly, the encrypted model was confirmed to be robust against a random key attack.

\begin{table}[t]
    \centering
    \caption{Robustness against use of plain images}
    \begin{tabular}{c|cc}
    \hline
    & \multicolumn{2}{c}{Test Image} \\
    \cline{2-3}
    Model & Plain & Ours \\
    \hline
    Baseline & \textbf{99.03} & - \\
    Ours & 9.06 & \textbf{99.03} \\
    \hline
    \end{tabular}
    \label{tab:2}
\end{table}

\begin{figure}[t]
    \centering
    \includegraphics[scale=0.5]{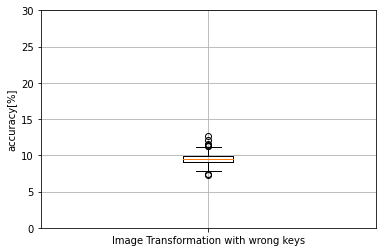}
    \caption{Evaluating robustness against random key attack. Boxes span from first to third quartile, referred to as $Q1$ and $Q3$, and whiskers show maximum and minimum values in range of $[Q1 - 1.5(Q3 - Q1), Q3 + 1.5(Q3 - Q1)]$. Band inside box indicates median. Outliers are indicated as dots.}
    \label{fig:6}
\end{figure}

\subsection{Combined Use of Federated Learning and Image Encryption}
The effectiveness of encrypted ViT models was finally evaluated under the combined use of federated learning (FL) and image encryption as shown Figure 5 \cite{nagamori2023combined}.

\subsubsection{Setup}
Experiments were conducted on the CIFAR-10 and CIFAR-100 datasets, where images were resized from $3 \times 32 \times 32$ to $3 \times 224 \times 224$ because we used ViT pre-trained with ImageNet-1K as a model.
For training models with FL, 10 clients were assumed, where each client had 5, 000 training images and 1, 000 test images.
Also, we used FedAVG \cite{McMahan2017communication} as the method of model integration.
Models were trained using stochastic gradient descent (SGD) with an initial learning
rate of $10^{-3}$, a momentum of $0.9$, and a batch size of 8.
We also used the cross-entropy loss function.
In addition, models were integrated every epoch, and the total number of epochs was set to $10$.
After the tenth integration, the integrated model was encrypted with secret keys, and every client used the secret keys to encrypt their test images.

\subsubsection{Classification Performance}
We evaluated the performance of the models in terms of classification accuracy.
Table 3 shows the experimental results for the CIFAR-10 and CIFAR-100 datasets, which  have 10 and 100 classes, respectively.
``Integrated  Model" indicates the results when the encrypted test images were applied to encrypted integrated models, and ``Baseline" represents the results when plain test images were applied to the plain models normally trained with plain images. \par
From the results, the combined use of FL and encrypted images was verified to have the same accuracy as that of models normally trained with plain images. Accordingly, the method in Figure 6 allows us not only to train models over multiple participants without directly sharing raw data but to also protect the visual information of test images.
In addition, our method enables us to easily update the key without re-training models, so each user can use an independent key to protect test images and their model.

\begin{table}[t]
    \centering
    \caption{Classification accuracy of proposed method}
    \begin{tabular}{c|cc}
        \hline
        & Integrated Model & Baseline \\
        \hline
        CIFAR-10 & 97.7 & 97.8 \\
        \hline
        CIFAR-100 & 85.1 & 85.1 \\
        \hline
    \end{tabular}
    \label{tab:3}
\end{table}

\section{Conclusion}
In this paper, we presented a block-wise encryption method for ViT and its applications.
The presented framework presented with the encryption method was verified to provide the same performance as that without any encryption since the embedding structure of ViT has a high similarity to block-wise encryption.
In addition, three applications of the method: privacy-preserving image classification, access control, and the combined use of federated learning and the encryption were conducted to show the effectiveness of the method for reliable DNNs.
In experiments, the method was demonstrated to outperform state-of-the-art methods with conventional methods for image encryption in terms of classification accuracy, and it was also verified to be effective in terms of the reliability of DNN models.

% \appendices

% % you can choose not to have a title for an appendix
% % if you want by leaving the argument blank
% \section{}

%References



% \end{thebibliography}

\bibliographystyle{ieicetr}% bib style
\bibliography{main}% your bib database

\begin{thebibliography}{10}
\providecommand{\url}[1]{#1}
\csname url@samestyle\endcsname
\providecommand{\newblock}{\relax}
\providecommand{\bibinfo}[2]{#2}
\providecommand{\BIBentrySTDinterwordspacing}{\spaceskip=0pt\relax}
\providecommand{\BIBentryALTinterwordstretchfactor}{4}
\providecommand{\BIBentryALTinterwordspacing}{\spaceskip=\fontdimen2\font plus
\BIBentryALTinterwordstretchfactor\fontdimen3\font minus
  \fontdimen4\font\relax}
\providecommand{\BIBforeignlanguage}[2]{{%
\expandafter\ifx\csname l@#1\endcsname\relax
\typeout{** WARNING: IEEEtran.bst: No hyphenation pattern has been}%
\typeout{** loaded for the language `#1'. Using the pattern for}%
\typeout{** the default language instead.}%
\else
\language=\csname l@#1\endcsname
\fi
#2}}
\providecommand{\BIBdecl}{\relax}
\BIBdecl

\bibitem{lecun2015deep}
Y.~LeCun, Y.~Bengio, and G.~Hinton, ``Deep learning,'' \emph{nature}, vol. 521,
  no. 7553, p. 436, 2015.

\bibitem{liu2019recent}
X.~Liu, Z.~Deng, and Y.~Yang, ``Recent progress in semantic image
  segmentation,'' \emph{Artif. Intell. Rev.}, vol.~52, no.~2, pp. 1089--1106,
  2019.

\bibitem{huang2014survey}
C.-T. Huang, L.~Huang, Z.~Qin, H.~Yuan, L.~Zhou, V.~Varadharajan, and C.-C.~J.
  Kuo, ``Survey on securing data storage in the cloud,'' \emph{APSIPA
  Transactions on Signal and Information Processing}, vol.~3, p.~e7, 2014.

\bibitem{moo2013p3}
M.-R. Ra, R.~Govindan, and A.~Ortega, ``P3: Toward {Privacy-Preserving} photo
  sharing,'' in \emph{10th USENIX Symposium on Networked Systems Design and
  Implementation (NSDI 13)}.\hskip 1em plus 0.5em minus 0.4em\relax Lombard,
  IL: USENIX Association, Apr. 2013, pp. 515--528.

\bibitem{lagendijk2013encrypted}
R.~Lagendijk, Z.~Erkin, and M.~Barni, ``Encrypted signal processing for privacy
  protection: Conveying the utility of homomorphic encryption and multiparty
  computation,'' \emph{IEEE Signal Processing Magazine}, vol.~30, no.~1, pp.
  82--105, 2013.

\bibitem{fredrikson2015model}
M.~Fredrikson, S.~Jha, and T.~Ristenpart, ``Model inversion attacks that
  exploit confidence information and basic countermeasures,'' in
  \emph{Proceedings of the 22nd ACM SIGSAC Conference on Computer and
  Communications Security}, ser. CCS '15.\hskip 1em plus 0.5em minus
  0.4em\relax New York, NY, USA: Association for Computing Machinery, 2015, pp.
  1322--1333.

\bibitem{shokri2017membership}
R.~Shokri, M.~Stronati, C.~Song, and V.~Shmatikov, ``Membership inference
  attacks against machine learning models,'' in \emph{2017 IEEE symposium on
  security and privacy (SP)}.\hskip 1em plus 0.5em minus 0.4em\relax IEEE,
  2017, pp. 3--18.

\bibitem{christian2014intriguing}
C.~Szegedy, W.~Zaremba, I.~Sutskever, J.~Bruna, D.~Erhan, I.~J. Goodfellow, and
  R.~Fergus, ``Intriguing properties of neural networks,'' in \emph{2nd
  International Conference on Learning Representations, {ICLR} 2014, Banff, AB,
  Canada, April 14-16, 2014, Conference Track Proceedings}, Y.~Bengio and
  Y.~LeCun, Eds., 2014.

\bibitem{siva2020adversarial}
R.~S. Siva~Kumar, M.~Nyström, J.~Lambert, A.~Marshall, M.~Goertzel,
  A.~Comissoneru, M.~Swann, and S.~Xia, ``Adversarial machine learning-industry
  perspectives,'' in \emph{2020 IEEE Security and Privacy Workshops (SPW)},
  2020, pp. 69--75.

\bibitem{kiya2022overview}
\BIBentryALTinterwordspacing
H.~Kiya, A.~P.~M. Maung, Y.~Kinoshita, S.~Imaizumi, and S.~Shiota, ``An
  overview of compressible and learnable image transformation with secret key
  and its applications,'' \emph{APSIPA Transactions on Signal and Information
  Processing}, vol.~11, no. 1, e11, 2022. [Online]. Available:
  \url{http://dx.doi.org/10.1561/116.00000048}
\BIBentrySTDinterwordspacing

\bibitem{Alexey2021an}
A.~Dosovitskiy, L.~Beyer, A.~Kolesnikov, D.~Weissenborn, X.~Zhai,
  T.~Unterthiner, M.~Dehghani, M.~Minderer, G.~Heigold, S.~Gelly, J.~Uszkoreit,
  and N.~Houlsby, ``An image is worth 16x16 words: Transformers for image
  recognition at scale,'' in \emph{International Conference on Learning
  Representations}, 2021.

\bibitem{chuman2019encryption}
T.~Chuman, W.~Sirichotedumrong, and H.~Kiya, ``Encryption-then-compression
  systems using grayscale-based image encryption for jpeg images,'' \emph{IEEE
  Transactions on Information Forensics and Security}, vol.~14, no.~6, pp.
  1515--1525, 2019.

\bibitem{chuman2017security}
T.~Chuman, K.~Kurihara, and H.~Kiya, ``On the security of block
  scrambling-based etc systems against jigsaw puzzle solver attacks,'' in
  \emph{2017 IEEE International Conference on Acoustics, Speech and Signal
  Processing (ICASSP)}, 2017, pp. 2157--2161.

\bibitem{zhou2014designing}
J.~Zhou, X.~Liu, O.~C. Au, and Y.~Y. Tang, ``Designing an efficient image
  encryption-then-compression system via prediction error clustering and random
  permutation,'' \emph{IEEE Transactions on Information Forensics and
  Security}, vol.~9, no.~1, pp. 39--50, 2014.

\bibitem{ghonge2014a}
M.~Ghonge and K.~Nimbokar, ``A survey based on designing an efficient image
  encryption-then-compression system,'' \emph{International Journal of Computer
  Applications}, p. 8887, 2014.

\bibitem{liu2018ecg}
T.~Y. Liu, K.~J. Lin, and H.~C. Wu, ``Ecg data encryption then compression
  using singular value decomposition,'' \emph{IEEE Journal of Biomedical and
  Health Informatics}, vol.~22, no.~3, pp. 707--713, 2018.

\bibitem{liu2010efficient}
W.~Liu, W.~Zeng, L.~Dong, and Q.~Yao, ``Efficient compression of encrypted
  grayscale images,'' \emph{IEEE Transactions on Image Processing}, vol.~19,
  no.~4, pp. 1097--1102, 2010.

\bibitem{hu2014a}
R.~Hu, X.~Li, and B.~Yang, ``A new lossy compression scheme for encrypted
  gray-scale images,'' in \emph{2014 IEEE International Conference on
  Acoustics, Speech and Signal Processing (ICASSP)}, 2014, pp. 7387--7390.

\bibitem{johnson2004on}
M.~Johnson, P.~Ishwar, V.~Prabhakaran, D.~Schonberg, and K.~Ramchandran, ``On
  compressing encrypted data,'' \emph{IEEE Transactions on Signal Processing},
  vol.~52, no.~10, pp. 2992--3006, 2004.

\bibitem{methaq2016an}
M.~Gaata and F.~F. Hantoosh, ``An efficient image encryption technique using
  chaotic logistic map and rc4 stream cipher,'' \emph{International Journal of
  Modern Trends in Engineering and Research}, vol.~3, pp. 213--218, 2016.

\bibitem{grayscale2019warit}
W.~Sirichotedumrong and H.~Kiya, ``Grayscale-based block scrambling image
  encryption using {YCbCr} color space for encryption-then-compression
  systems,'' \emph{APSIPA Transactions on Signal and Information Processing},
  vol.~8, p.~e7, 2019.

\bibitem{Shoko2020a}
S.~Imaizumi, Y.~Izawa, R.~Hirasawa, and H.~Kiya, ``A reversible data hiding
  method in compressible encrypted images,'' \emph{IEICE Transactions on
  Fundamentals of Electronics, Communications and Computer Sciences}, vol.
  E103.A, no.~12, pp. 1579--1588, 2020.

\bibitem{iida2020privacy}
K.~Iida and H.~Kiya, ``Privacy-preserving content-based image retrieval using
  compressible encrypted images,'' \emph{IEEE Access}, vol.~8, pp.
  200\,038--200\,050, 2020.

\bibitem{iida2019an}
------, ``An image identification scheme of encrypted jpeg images for
  privacy-preserving photo sharing services,'' \emph{2019 IEEE International
  Conference on Image Processing (ICIP)}, pp. 4564--4568, 2019.

\bibitem{kawamura2020aprivacy}
A.~Kawamura, Y.~Kinoshita, T.~Nakachi, S.~Shiota, and H.~Kiya, ``A
  privacy-preserving machine learning scheme using etc images,'' \emph{IEICE
  Transactions on Fundamentals of Electronics, Communications and Computer
  Sciences}, vol. E103.A, no.~12, pp. 1571--1578, 2020.

\bibitem{bandoh2020distributed}
Y.~Bandoh, T.~Nakachi, and H.~Kiya, ``Distributed secure sparse modeling based
  on random unitary transform,'' \emph{IEEE Access}, vol.~8, pp.
  211\,762--211\,772, 2020.

\bibitem{takayuki2020secure}
T.~Nakachi, Y.~Bandoh, and H.~Kiya, ``Secure overcomplete dictionary learning
  for sparse representation,'' \emph{IEICE Transactions on Information and
  Systems}, vol. E103.D, no.~1, pp. 50--58, 2020.

\bibitem{nakachi2020privacy}
T.~Nakachi, Y.~Wang, and H.~Kiya, ``Privacy-preserving pattern recognition
  using encrypted sparse representations in l0 norm minimization,'' in
  \emph{ICASSP 2020 - 2020 IEEE International Conference on Acoustics, Speech
  and Signal Processing (ICASSP)}, 2020, pp. 2697--2701.

\bibitem{ibuki2016unitary}
I.~Nakamura, Y.~Tonomura, and H.~Kiya, ``Unitary transform-based template
  protection and its application to {$l^{2}$}-norm minimization problems,''
  \emph{IEICE Transactions on Information and Systems}, vol. E99.D, no.~1, pp.
  60--68, 2016.

\bibitem{maekawa2019privacy}
T.~Maekawa, A.~Kwamura, T.~Nakachi, and H.~Kiya, ``Privacy-preserving support
  vector machine computing using random unitary transformation,'' \emph{IEICE
  Transactions on Fundamentals of Electronics, Communications and Computer
  Sciences}, vol. E102.A, no.~12, pp. 1849--1855, 2019.

\bibitem{aprilpyone2021block}
M.~Aprilpyone and H.~Kiya, ``Block-wise image transformation with secret key
  for adversarially robust defense,'' \emph{IEEE Transactions on Information
  Forensics and Security}, vol.~16, pp. 2709--2723, 2021.

\bibitem{maung2020encryption}
------, ``Encryption inspired adversarial defense for visual classification,''
  in \emph{2020 IEEE International Conference on Image Processing (ICIP)},
  2020, pp. 1681--1685.

\bibitem{maung2021ensemble}
------, ``Ensemble of key-based models: Defense against black-box adversarial
  attacks,'' in \emph{2021 IEEE 10th Global Conference on Consumer Electronics
  (GCCE)}, 2021, pp. 95--98.

\bibitem{chen2018protect}
M.~Chen and M.~Wu, ``Protect your deep neural networks from piracy,'' in
  \emph{2018 IEEE International Workshop on Information Forensics and Security
  (WIFS)}, 2018, pp. 1--7.

\bibitem{chen2019deepattest}
H.~Chen, C.~Fu, B.~D. Rouhani, J.~Zhao, and F.~Koushanfar, ``Deepattest: An
  end-to-end attestation framework for deep neural networks,'' in \emph{2019
  ACM/IEEE 46th Annual International Symposium on Computer Architecture
  (ISCA)}, 2019, pp. 487--498.

\bibitem{maungmaung2021a}
M.~Aprilpyone and H.~Kiya, ``A protection method of trained cnn model with a
  secret key from unauthorized access,'' \emph{APSIPA Transactions on Signal
  and Information Processing}, vol.~10, p. e10, 2021.

\bibitem{uchida2017embedding}
Y.~Uchida, Y.~Nagai, S.~Sakazawa, and S.~Satoh, ``Embedding watermarks into
  deep neural networks,'' in \emph{Proceedings of the 2017 ACM on International
  Conference on Multimedia Retrieval}, ser. ICMR '17.\hskip 1em plus 0.5em
  minus 0.4em\relax Association for Computing Machinery, 2017, pp. 269--277.

\bibitem{chen2019deepmarks}
H.~Chen, B.~D. Rouhani, C.~Fu, J.~Zhao, and F.~Koushanfar, ``Deepmarks: A
  secure fingerprinting framework for digital rights management of deep
  learning models,'' in \emph{Proceedings of the 2019 on International
  Conference on Multimedia Retrieval}, ser. ICMR '19.\hskip 1em plus 0.5em
  minus 0.4em\relax Association for Computing Machinery, 2019, p. 105–113.

\bibitem{rouhani2018deepsigns}
\BIBentryALTinterwordspacing
B.~D. Rouhani, H.~Chen, and F.~Koushanfar, ``Deepsigns: A generic watermarking
  framework for ip protection of deep learning models,'' \emph{arXiv preprint
  arXiv:1804.00750}, 2018. [Online]. Available:
  \url{https://arxiv.org/abs/1804.00750}
\BIBentrySTDinterwordspacing

\bibitem{fan2021deepip}
L.~Fan, K.~W. Ng, C.~S. Chan, and Q.~Yang, ``Deepip: Deep neural network
  intellectual property protection with passports,'' \emph{IEEE Transactions on
  Pattern Analysis and Machine Intelligence}, 2021.

\bibitem{adi2018turning}
Y.~Adi, C.~Baum, M.~Cisse, B.~Pinkas, and J.~Keshet, ``Turning your weakness
  into a strength: Watermarking deep neural networks by backdooring,'' in
  \emph{Proceedings of the 27th USENIX Conference on Security Symposium}, ser.
  SEC'18.\hskip 1em plus 0.5em minus 0.4em\relax USENIX Association, 2018, pp.
  1615--1631.

\bibitem{zhang2018protecting}
J.~Zhang, Z.~Gu, J.~Jang, H.~Wu, M.~P. Stoecklin, H.~Huang, and I.~Molloy,
  ``Protecting intellectual property of deep neural networks with
  watermarking,'' in \emph{Proceedings of the 2018 on Asia Conference on
  Computer and Communications Security}, ser. ASIACCS '18.\hskip 1em plus 0.5em
  minus 0.4em\relax Association for Computing Machinery, 2018, pp. 159--172.

\bibitem{sakazawa2019visual}
S.~Sakazawa, E.~Myodo, K.~Tasaka, and H.~Yanagihara, ``Visual decoding of
  hidden watermark in trained deep neural network,'' in \emph{2019 IEEE
  Conference on Multimedia Information Processing and Retrieval (MIPR)}, 2019,
  pp. 371--374.

\bibitem{le2019adversarial}
E.~Le~Merrer, P.~P{\'e}rez, and G.~Tr{\'e}dan, ``{Adversarial frontier
  stitching for remote neural network watermarking},'' \emph{{Neural Computing
  and Applications}}, vol.~32, no.~13, pp. 9233--9244, 2019.

\bibitem{devlin2019bert}
J.~Devlin, M.-W. Chang, K.~Lee, and K.~Toutanova, ``{BERT}: Pre-training of
  deep bidirectional transformers for language understanding,'' in
  \emph{Proceedings of the 2019 Conference of the North {A}merican Chapter of
  the Association for Computational Linguistics: Human Language Technologies,
  Volume 1 (Long and Short Papers)}.\hskip 1em plus 0.5em minus 0.4em\relax
  Association for Computational Linguistics, Jun. 2019, pp. 4171--4186.

\bibitem{ilya2021mlpmixer}
I.~Tolstikhin, N.~Houlsby, A.~Kolesnikov, L.~Beyer, X.~Zhai, T.~Unterthiner,
  J.~Yung, A.~P. Steiner, D.~Keysers, J.~Uszkoreit, M.~Lucic, and
  A.~Dosovitskiy, ``{MLP}-mixer: An all-{MLP} architecture for vision,'' in
  \emph{Advances in Neural Information Processing Systems}, A.~Beygelzimer,
  Y.~Dauphin, P.~Liang, and J.~W. Vaughan, Eds., 2021.

\bibitem{hugo2021resmlp}
H.~Touvron, P.~B. andMathilde Caron, M.~Cord, A.~El{-}Nouby, E.~Grave,
  A.~Joulin, G.~Synnaeve, J.~Verbeek, and H.~J{\'{e}}gou, ``Resmlp: Feedforward
  networks for image classification with data-efficient training,''
  \emph{CoRR}, vol. abs/2105.03404, 2021.

\bibitem{chen2022cyclemlp}
S.~Chen, E.~Xie, C.~GE, R.~Chen, D.~Liang, and P.~Luo, ``Cycle{MLP}: A
  {MLP}-like architecture for dense prediction,'' in \emph{International
  Conference on Learning Representations}, 2022.

\bibitem{liu2021pay}
H.~Liu, Z.~Dai, D.~So, and Q.~V. Le, ``Pay attention to {MLP}s,'' in
  \emph{Advances in Neural Information Processing Systems}, A.~Beygelzimer,
  Y.~Dauphin, P.~Liang, and J.~W. Vaughan, Eds., 2021.

\bibitem{Hou2022VisionPA}
Q.~Hou, Z.~Jiang, L.~Yuan, M.-M. Cheng, S.~Yan, and J.~Feng, ``Vision
  permutator: A permutable mlp-like architecture for visual recognition,''
  \emph{IEEE transactions on pattern analysis and machine intelligence},
  vol.~PP, 2022.

\bibitem{trockman2022patches}
\BIBentryALTinterwordspacing
A.~Trockman and J.~Z. Kolter, ``Patches are all you need?'' \emph{arXiv
  preprint arXiv:2201.09792}, 2022. [Online]. Available:
  \url{https://arxiv.org/abs/2201.09792}
\BIBentrySTDinterwordspacing

\bibitem{wang2007face}
Y.~Wang and K.~Plataniotis, ``Face based biometric authentication with
  changeable and privacy preservable templates,'' in \emph{2007 Biometrics
  Symposium}, 2007, pp. 1--6.

\bibitem{hitoshi2023image}
H.~Kiya, R.~Iijima, A.~MaungMaung, and Y.~Kinoshita, ``Image and model
  transformation with secret key for vision transformer,'' \emph{IEICE
  Transactions on Information and Systems}, vol. E106.D, no.~1, pp. 2--11,
  2023.

\bibitem{kiya2022privacy}
\BIBentryALTinterwordspacing
H.~Kiya, T.~Nagamori, S.~Imaizumi, and S.~Shiota, ``Privacy-preserving semantic
  segmentation using vision transformer,'' \emph{Journal of Imaging}, vol.~8,
  no.~9, 2022. [Online]. Available:
  \url{https://www.mdpi.com/2313-433X/8/9/233}
\BIBentrySTDinterwordspacing

\bibitem{jakub2016federated}
\BIBentryALTinterwordspacing
J.~Konečný, H.~B. McMahan, F.~X. Yu, P.~Richtarik, A.~T. Suresh, and
  D.~Bacon, ``Federated learning: Strategies for improving communication
  efficiency,'' in \emph{NIPS Workshop on Private Multi-Party Machine
  Learning}, 2016. [Online]. Available: \url{https://arxiv.org/abs/1610.05492}
\BIBentrySTDinterwordspacing

\bibitem{McMahan2017communication}
B.~McMahan, E.~Moore, D.~Ramage, S.~Hampson, and B.~A.~y. Arcas,
  ``{Communication-Efficient Learning of Deep Networks from Decentralized
  Data},'' in \emph{Proceedings of the 20th International Conference on
  Artificial Intelligence and Statistics}, ser. Proceedings of Machine Learning
  Research, A.~Singh and J.~Zhu, Eds., vol.~54.\hskip 1em plus 0.5em minus
  0.4em\relax PMLR, 20--22 Apr 2017, pp. 1273--1282.

\bibitem{paszke2019pytorch}
A.~Paszke, S.~Gross, F.~Massa, A.~Lerer, J.~Bradbury, G.~Chanan, T.~Killeen,
  Z.~Lin, N.~Gimelshein, L.~Antiga, A.~Desmaison, A.~Kopf, E.~Yang, Z.~DeVito,
  M.~Raison, A.~Tejani, S.~Chilamkurthy, B.~Steiner, L.~Fang, J.~Bai, and
  S.~Chintala, ``Pytorch: An imperative style, high-performance deep learning
  library,'' in \emph{Advances in Neural Information Processing Systems 32},
  H.~Wallach, H.~Larochelle, A.~Beygelzimer, F.~d\textquotesingle
  Alch\'{e}-Buc, E.~Fox, and R.~Garnett, Eds.\hskip 1em plus 0.5em minus
  0.4em\relax Curran Associates, Inc., 2019, pp. 8024--8035.

\bibitem{tanaka2018learnable}
M.~Tanaka, ``Learnable image encryption,'' in \emph{2018 IEEE International
  Conference on Consumer Electronics-Taiwan (ICCE-TW)}, 2018, pp. 1--2.

\bibitem{sirichotedumrong2019privacy}
W.~Sirichotedumrong, T.~Maekawa, Y.~Kinoshita, and H.~Kiya,
  ``Privacy-preserving deep neural networks with pixel-based image encryption
  considering data augmentation in the encrypted domain,'' in \emph{2019 IEEE
  International Conference on Image Processing (ICIP)}, 2019, pp. 674--678.

\bibitem{sirichotedumrong2021a}
W.~Sirichotedumrong and H.~Kiya, ``A gan-based image transformation scheme for
  privacy-preserving deep neural networks,'' in \emph{2020 28th European Signal
  Processing Conference (EUSIPCO)}, 2021, pp. 745--749.

\bibitem{ito2021image}
H.~Ito, Y.~Kinoshita, M.~Aprilpyone, and H.~Kiya, ``Image to perturbation: An
  image transformation network for generating visually protected images for
  privacy-preserving deep neural networks,'' \emph{IEEE Access}, vol.~9, pp.
  64\,629--64\,638, 2021.

\bibitem{chuman2023ajigsaw}
\BIBentryALTinterwordspacing
T.~Chuman and H.~Kiya, ``A jigsaw puzzle solver-based attack on image
  encryption using vision transformer for privacy-preserving dnns,''
  \emph{Information}, vol.~14, no.~6, 2023. [Online]. Available:
  \url{https://www.mdpi.com/2078-2489/14/6/311}
\BIBentrySTDinterwordspacing

\bibitem{nagamori2023combined}
T.~Nagamori and H.~Kiya, ``Combined use of federated learning and image
  encryption for privacy-preserving image classification with vision
  transformer,'' \emph{RISP International Workshop on Nonlinear Circuits,
  Communications and Signal Processing,}, 2023.

\end{thebibliography}

\end{document}